\documentclass[aps,prl,10pt,twocolumn,amsmath,amssymb,longbibliography,superscriptaddress]{revtex4-1}
\usepackage{graphicx,xcolor}
\makeatletter
\def\amsbb{\use@mathgroup \M@U \symAMSb}
\makeatother
\usepackage{bbm,overpic,mathrsfs}
\usepackage{amsmath,amsfonts}
\makeatletter
\def\amsbb{\use@mathgroup \M@U \symAMSb}
\makeatother
\usepackage[bbgreekl]{mathbbol}
\usepackage{siunitx}
\newcommand\ve[1]{\boldsymbol{#1}}
\newcommand{\ma}[1]{\ensuremath{\amsbb{#1}}}

\begin{document}
\title{Transport barriers for microswimmers in unsteady flow}
\author{L. Storm}
\affiliation{Department of Physics, University of Gothenburg, 41296 Gothenburg, Sweden}
\author{J. Qiu}
\affiliation{Department of Physics, University of Gothenburg, 41296 Gothenburg, Sweden}
\author{K. Gustavsson}
\affiliation{Department of Physics, University of Gothenburg, 41296 Gothenburg, Sweden}
\author{B. Mehlig}
\affiliation{Department of Physics, University of Gothenburg, 41296 Gothenburg, Sweden}

\begin{abstract}
Certain flow structures trap active microorganisms over extended periods, impacting mortality and growth by inhibiting foraging, exploration, and predator evasion. We show that transport barriers and trapping of a microswimmer in unsteady vortex flow can be understood in terms of Lagrangian coherent  structures (LCS) in phase space,  taking into account both flow and swimming dynamics. These phase-space LCS constrain when and how artificial microswimmers can be trained by reinforcement learning to escape
traps.
Our results suggest that phase-space LCS can explain training failure and success 
in a broad class of navigation problems. 
\end{abstract}

\maketitle

Transport barriers for microswimmers
are structures that restrict exchange between different regions~\cite{wiggins2005the}, playing a crucial role in limiting microorganism transport in the ocean~\cite{levy2018submesoscale,prants2023transport}, in microfluidic devices~\cite{rusconi2014bacterial,ran2024enhancing}, and bioreactors~\cite{lara2006living}.
Experiments
\cite{kessler1985hydrodynamic,durham2009disruption,durham2013turbulence,rusconi2014bacterial,delillo2014turbulent,mathijssen2019oscillatory,dehkharghani2019bacterial,si2021preferential,berman2021transport,qin2022confinement,ran2024enhancing},
numerical simulations
\cite{torney2007transport,durham2011gyrotactic,khurana2011reduced,durham2013turbulence,zhan2013accumulation,delillo2014turbulent,borgnino2019alignment,borgnino2022alignment},
and analytical studies
\cite{gustavsson2016preferential,borgnino2019alignment,arguedasleiva2020microswimmers,borgnino2022alignment,tanasijevic2022microswimmers} show that the interplay of swimming, rotational dynamics, and flow structures can lead to accumulation in certain locations~\cite{durham2013turbulence,zhan2013accumulation,gustavsson2016preferential,dehkharghani2019bacterial,mathijssen2019oscillatory}, preferential alignment~\cite{borgnino2019alignment,mathijssen2019oscillatory,borgnino2022alignment}, and trapping in Poiseuille flow~\cite{kessler1985hydrodynamic,zoettl2012nonlinear}, shear layers~\cite{durham2009disruption,rusconi2014bacterial}, or vortex cores~\cite{torney2007transport,durham2011gyrotactic,khurana2011reduced,delillo2014turbulent,arguedasleiva2020microswimmers,qin2022confinement,tanasijevic2022microswimmers}.
These mechanisms alter encounter rates, hinder foraging, and reduce mixing, potentially increasing mortality.
How  prey can escape predator feeding currents is particularly important. However, it remains unclear to which extent microorganisms can avoid or escape barriers and traps through active flow sensing or passive strategies such as optimised run-and-tumble dynamics.

In steady flow,  
saddle manifolds~\cite{arguedasleiva2020microswimmers,berman2021transport,le2024barriers} and closed orbits of the flow~\cite{tanasijevic2022microswimmers} form transport barriers. 
In unsteady flow -- the generic case -- 
barriers for microswimmers are less well understood. Saddle manifolds and closed orbits generalise to hyperbolic and elliptic Lagrangian coherent structures (LCS)~\cite{haller2000lagrangian,shadden2005definition,haller2015lagrangian,oettinger2016global}.
Microswimmer accumulation and trapping 
~\cite{khurana2012interactions,si2021preferential}, optimal navigation   with instantaneous steering~\cite{krishna2023finite}, and 
clustering of elongated swimmers~\cite{ran2024enhancing}
can be linked to the flow LCS. 

However, LCS of the fluid flow alone cannot determine whether a region is a trap for a microswimmer, whether escape is possible, and by what strategy. The problem is that the LCS of the flow ignores swimming dynamics. Instead one must search for LCS in the entire phase space of the problem, including both fluid flow {\em and} swimming dynamics.
To illustrate the power of this phase-space approach, we analyse elliptic and hyperbolic LCS in phase space to identify traps and barriers for a swimmer in a two-dimensional vortex flow (our findings generalise to more complex flow structures).
Around each vortex, we find a family of nested elliptic LCS in the form of extended phase-space tubes that resist deformation.
Each of these structures traps the swimmer, confining it to quasi-periodic motion, while the dynamics is chaotic outside the outermost  elliptic LCS (the \lq{}Lagrangian vortex boundary\rq{}~\cite{blazevski2014hyperbolic}). 
Even outside, escape can be slow, when hyperbolic LCS form intricate folded phase-space barriers that force swimmers along winding paths before leaving the vortex cell.
The nested LCS tilt  non-uniformly, more for larger swimming speeds, enabling to escape traps and navigate barriers by suitable turning.
Using stochastic reorientation, the swimmer can escape, but crossing the family of nested elliptic LCS tends to be slow, similar to escape by weak  isotropic diffusion in two-dimensional systems~\cite{haller2018material}. We find more efficient escape strategies by reinforcement learning,
guided by the phase-space
structure dictated by the LCS. 

This progress is important, because efficient microswimmer navigation is a complex problem. How can a swimmer exploit flow structures for speed, avoid phase-space barriers, and recognise how to adjust its strategy to escape? Recent work addresses this for tasks such as directional navigation~\cite{colabrese2017flow,qiu2022Navigation,xu2023long}, point-to-point navigation~\cite{biferale2019zermelo,schneider2019optimal,alageshan2020machine,gunnarson2021Learning}, clustering~\cite{calascibetta2023taming}, and targeting specific flow regions~\cite{mousavi2025short}, relying on heuristic signals with minimal prior information about the underlying dynamics. 
Our results illustrate how phase-space LCS  can inform the search for strategies, and indicate which signals to consider. 
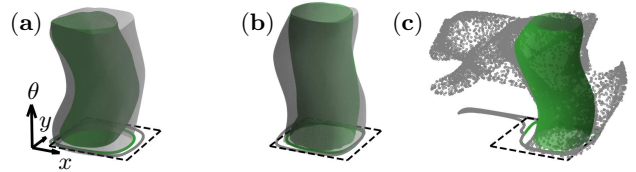
\begin{figure}
    \centering
    \begin{overpic}[width=0.45\textwidth]{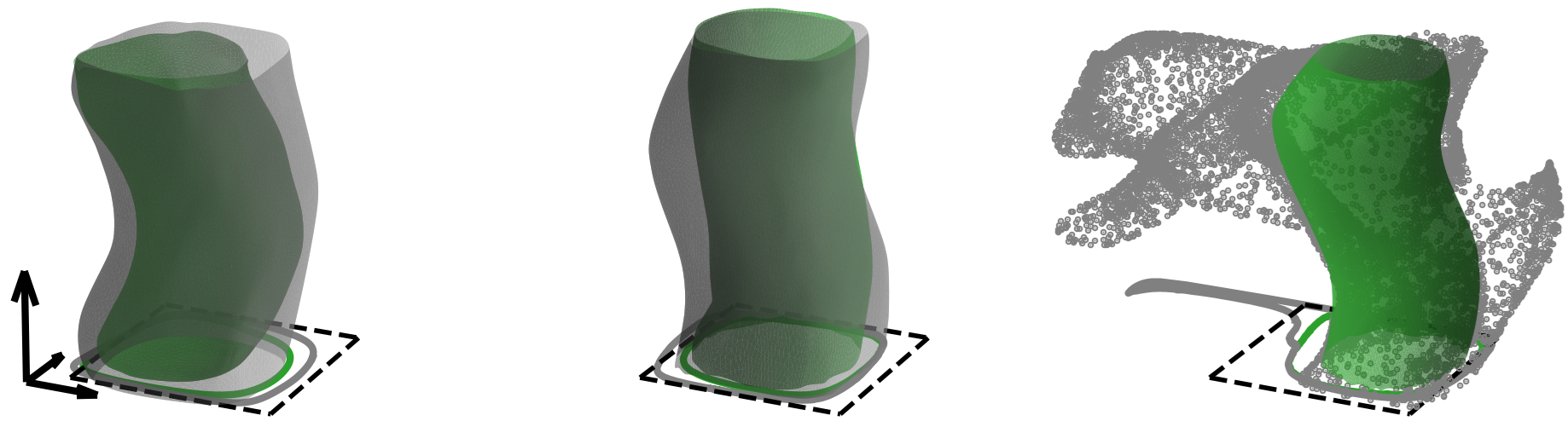}
    \put(-2,22){({\bf a})}
    \put(36,22){({\bf b})}
    \put(61,22){({\bf c})}
    \put(6,-1){$x$}
    \put(2.8,6){$y$}
    \put(0.9,10.5){$\theta$}
    \end{overpic}
    \caption{\label{fig:Fig1} Elliptic LCS  $\gamma_\tau(t)$ in an unsteady vortex flow (see text), for coherence times $\tau=5$ (grey) and $10$ (green) in phase space ($x$, $y$, $\theta$) for a single vortex cell at ({\bf a})  $t=0$,
    ({\bf b}) $t=5$, and ({\bf c}) $t=10$.
    Parameters:  $\Phi=0.1$, $\Xi=0$, $B=0.04$, $\omega\tau_{\rm K}=0.5$. Dashed lines show the  boundary of the flow vortex cell, solid lines show two-dimensional elliptic LCS  $\gamma^{({\rm flow})}_\tau(t)$ of the flow in the $x$-$y$-plane for $\tau=5$ (grey) and $\tau=10$ (green).
    }
\end{figure}

{\em Background}.
\label{sec:trapping}
In steady flow, transport barriers are commonly discussed in terms of invariant manifolds acting as separatrices in the phase-space of swimmer position and orientation~\cite{berman2021transport}.
For example, quasi-periodic phase-space manifolds  prevent mixing between interiour and exteriour trajectories~\cite{torney2007transport}. 
In unsteady flow, by contrast, there are no such long-term invariant structures acting as transport barriers, in general. Experiments \cite{qin2022confinement} and simulations \cite{torney2007transport,khurana2011reduced} indicate that swimmers in weakly time-periodic flow may remain trapped.
The trapping regions 
appear to be close to but not identical with the traps
of the unperturbed, time-independent flow. 
In particular, it is unclear
whether the traps persist ad infinitum, or whether the transport barriers are transient.

Lagrangian coherent structures (LCS) provide a natural framework for describing transient transport barriers and trapping in time-dependent flow. 
LCS are material surfaces 
with special properties.
Hyperbolic repelling LCS  -- the finite-time analogues of  
unstable saddle  manifolds~\cite{blazevski2014hyperbolic} -- are the locally most repelling material surfaces, located at ridges of large maximal finite-time Lypunov exponents (FTLE)~\cite{haller2011variational}. 
Elliptic LCS -- the finite-time analog of (quasi-)periodic invariant manifolds -- are closed material surfaces that experience uniform tangential stretching over a finite coherence time $\tau$, ensuring coherence by preventing anisotropic stretching and folding of the surface~\cite{oettinger2016global}.
Hyperbolic LCS form open finite-time transport barriers, while elliptic LCS enclose regions where particles remain trapped during $\tau$~\cite{haller2015lagrangian}.
In two-dimensional dynamical systems, elliptic LCS also
inhibit transport due to weak isotropic diffusion~\cite{haller2018material}.
We show that both types of LCS play important roles for  microswimmer transport in unsteady flow. 

{\em Model}.
We consider small spherical particle swimming  in the $x$-$y$-plane with swimming speed $v_{\rm s}$, 
swimming direction $\ve n=[\cos{\theta},\sin{\theta}]$,
and 
angular velocity~$\omega_{\rm s}$:
    \begin{align}
     \label{eq:swim_dyn}
        \dot{\ve r} = & \,\ve u+v_{\rm s}\ve n,\quad
        \dot{\theta} =  \,\tfrac{1}{2}(\nabla\wedge \ve u)\cdot\hat{\bf e}_z+\omega_{\rm s}\,.
    \end{align}
Here $\ve u(\ve r,t)$  is the externally imposed two-dimensional fluid flow with vorticity $\nabla\wedge \ve u$, $\ve r=[x,y]^{\sf T}$ is the location of the swimmer, 
and $\hat{\bf e}_z$
the unit vector normal to the $x$-$y$-plane. Dots denote derivaties w.r.t. time~$t$. We use $\ve u=\ve\nabla\psi\wedge\hat{\bf e}_z$ with stream function $\psi(\ve x,t)=u_0L_0\cos(x/L_0+2\pi B\sin{\omega t})\cos(y/L_0)$ with maximal velocity $u_0$, vortex size $\pi L_0$, and horizontal phase shift of amplitude $2\pi B$ and frequency $\omega$~\cite{solomon1988chaotic,camassa1991chaotic,torney2007transport}.
For $B=0$, one obtains the steady Taylor-Green vortex (TGV) flow~\cite{taylor1937mechanism}, for $B>0$ the flow becomes chaotic~\cite{camassa1991chaotic,rom1990analytical}.
In what follows, we non-dimensionalise time by $\tau_{\rm K}=L_0/u_0$ and position by $L_0$. We define the non-dimensional swimming speed $\Phi\equiv v_{\rm s}/u_0$, turning rate $\Xi = \omega_{\rm s}\tau_{\rm K}$, and set $\omega\tau_{\rm K}=0.5$ throughout. 
We consider the symmetry-reduced region $-\pi/2 \leq x,y \leq \pi /2$ in the $x$-$y$-plane (\lq vortex cell\rq{}).

{\em LCS for a swimmer in vortex flow}.
For constant $v_{\rm s}$ and $\omega_{\rm s}$, Eq.~(\ref{eq:swim_dyn}) conserves phase-space volume, leading to  
nested elliptic and hyperbolic LCS that stretch or compress locally without local volume change~\cite{haller2013coherent}.
We denote the 
outermost elliptic phase-space  LCS 
(the Lagrangian vortex boundary in phase space) evaluated at time $t$ by $\gamma_\tau(t)$. Here $\tau$ is the specified coherence time:  $\gamma_\tau(t)$ does not deform significantly for $0\leq t\leq\tau$, but stretches and folds for  longer times (see supplementary material (SM) for details~\cite{sm}). Figure~\ref{fig:Fig1} illustrates that LCS are inherently time-dependent, 
unlike invariant manifolds in steady flow.
 Panel~({\bf a}) shows the initial surfaces $\gamma_\tau(t=0)$ for $\tau=5$ (grey) and $\tau=10$ (green).
The surface $\gamma_{10}(0)$ lies inside $\gamma_{5}(0)$, because coherence decays under chaotic dynamics.
Panel~({\bf b}) shows that 
both $\gamma_5(t)$
and $\gamma_{10}(t)$ remain coherent for $t=5$, they essentially retain their shapes and thus confine the swimmers inside.
But the former does not stay coherent up to $t=10$ (its coherence time is only $\tau=5$), the material surface 
 is significantly stretched and folded [panel ({\bf c})]. By contrast, $\gamma_{10}(10)$ remains intact, 
so $\gamma_{10}(t)$  is a transport barrier up to $t=10$ because it prevents the swimmers inside from leaving the vortex cell during this time. Swimmers inside $\gamma_5(t)$, by contrast, 
explore phase space for $t>5$. In addition, while general material surfaces  are sensitive to small perturbations (passive-scalar mixing in turbulence \cite{shraiman2000scalar,dimotakis2005turbulent,villermaux2019mixing}), coherent LCS form robust transport barriers.

Let us contrast
the phase-space LCS $\gamma_\tau(t)$ with the two-dimensional LCS $\gamma_{\tau}^{({\rm flow})}(t)$ of the flow
~\cite{khurana2012interactions,si2021preferential,ran2024enhancing}.
The latter  are shown as solid lines in the $x$-$y$-plane in Fig.~\ref{fig:Fig1}.
First, $\gamma_\tau(t)$ enclose smaller volumes
than the $\theta$-independent tubes corresponding to $\gamma_{\tau}^{({\rm flow})}(t)$.  So swimming reduces the fraction of trapped swimmers.  Second, 
the  phase-space volume of $\gamma_\infty$ 
decreases as  $\Phi$ increases. For $B=0$,   the LCS disappears at $\Phi_c\approx0.416$ (Figure~S3 in the SM \cite{sm}).
This bifurcation explains why  traps in a TGV flow
disappear at a critical value of~$\Phi$~\cite{torney2007transport}.
Third, $\gamma_{\tau}^{({\rm flow})}$ misclassifies swimmers: those inside $\gamma_\tau^{({\rm flow})}$ but outside $\gamma_\tau$ may escape, while those outside $\gamma_\tau^{({\rm flow})}$ but inside $\gamma_\tau$ remain trapped.
Fourth, $\gamma_{\tau=5}^{({\rm flow})}$ predicts a single upper-left escape path for swimmers trapped at $t=5$ and escaping at $t=10$, while $\gamma_{\tau=5}(10)$ shows multiple $\theta$-dependent escape paths [Fig.~\ref{fig:Fig1}({\bf c})]. 
The tilt of $\gamma_\tau$ shows how swimmers may escape (or enter) the LCS via rapid reorientation, information entirely absent from~$\gamma_{\tau}^{({\rm flow})}$.

Next, we demonstrate that the elliptic LCS shown in Fig.~\ref{fig:Fig1} indeed act as transport barriers, and we discuss how hyperbolic LCS affect transport. 
Fig.~\ref{fig:Fig2}({\bf a}) shows 
the fraction $F_{\rm esc}(t)$ of swimmers that exited the vortex cell at least once by time $t$
(the swimmers are initially uniformly distributed in phase space).
\begin{figure}[t]
    \centering
    \begin{overpic}[width=\linewidth]{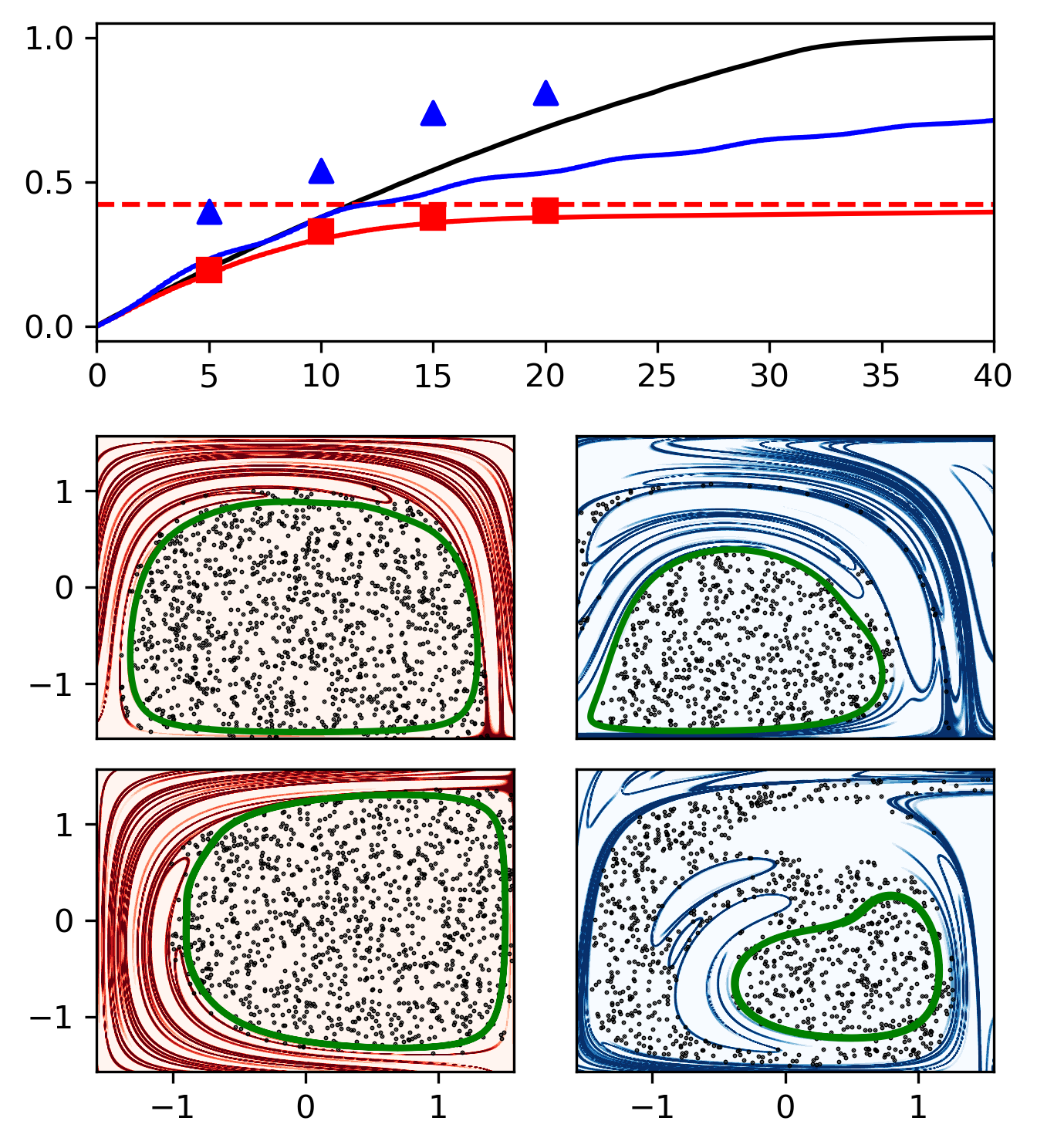}
    
    \put(1.5,89.5){$F_\text{esc}$}
    \put(81,66.5){$t$}
    \put(4.5,52){$y$}
    \put(4.5,23){$y$}
    \put(30.75,2.5){$x$}
    \put(72.65,2.5){$x$}

    \put(8.75,94){\colorbox{white}{({\bf a})}}
    \put(8.75,58){\colorbox{white}{({\bf b})}}
    \put(50.5,58){\colorbox{white}{({\bf d})}}
    \put(8.75,29.25){\colorbox{white}{({\bf c})}}
    \put(50.5,29.25){\colorbox{white}{({\bf e})}}
    \end{overpic}
    \caption{LCS as transport barriers. ({\bf a}) Fraction $F_{\rm esc}(t)$ of initially uniformly distributed swimmers that have escaped the TGV vortex cell at time $t$, for 
     steady flow, $B=0$ (solid red line ), unsteady flow, $B= 0.04$ (blue line), and for $\ve u=0$ (black line). The dashed red line shows unity minus the phase-space volume ratio of $\gamma_\infty$ and the vortex cell, explaining $F_{\rm esc}(\infty)$. 
    Markers show the corresponding expression  for $\gamma_t(t)$ for $B=0$ (red, $\square$) and $B=0.04$ (blue, $\vartriangle$). ({\bf b})-({\bf e}) Intersections of hyperbolic LCS (red \& blue  lines) and outermost elliptic LCS  $\gamma_{\tau=20}(0)$  (green) with  $\theta=0$ and $\theta=\pi/2$. 
    Black dots show the initial positions of swimmers that have not escaped the vortex cell until  $t=20$ (swimmers were initialised uniformly in the vortex cell).
    Parameters:  $\Phi=0.1$, $\Xi=0$, $\omega\tau_{\rm K}=0.5$. $B=0$ ({\bf b},{\bf c}),  and $B=0.04$~({\bf d},{\bf e}).
    }
    \label{fig:Fig2}
\end{figure}
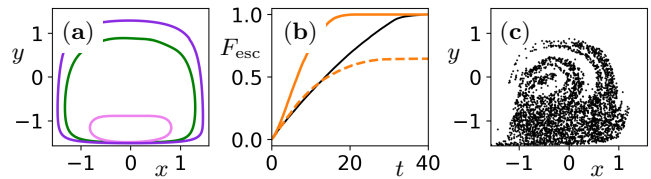
\begin{figure}[t]
    \centering
    \begin{overpic}[width=\linewidth]{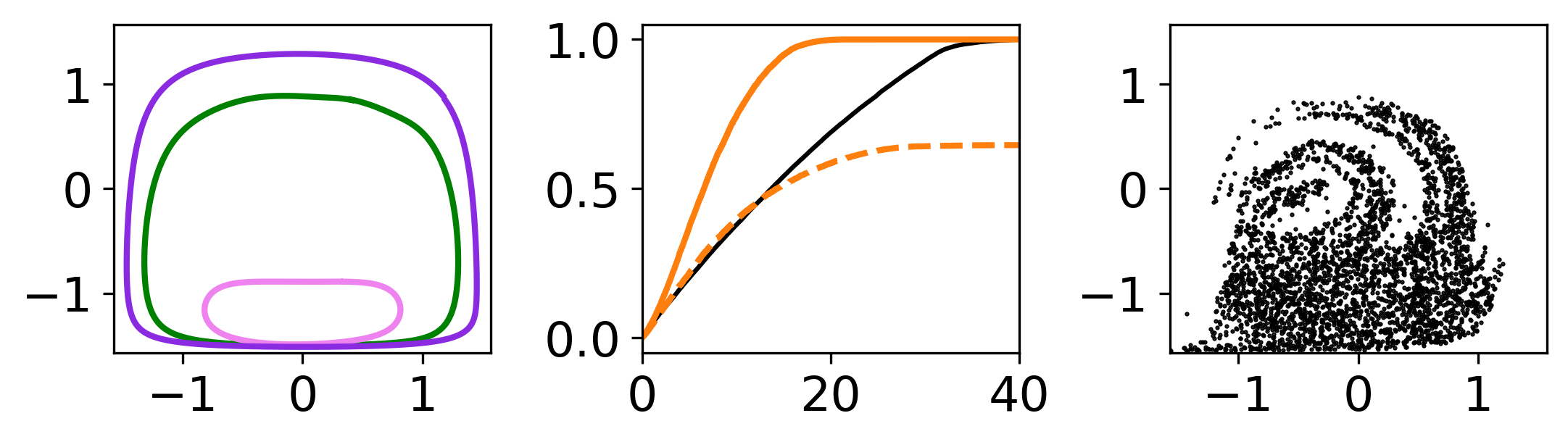}
    \put(7.5,22){\colorbox{white}{({\bf a})}}
     \put(41.5,22){\colorbox{white}{({\bf b})}}
      \put(76,22){({\bf c})}
    \put(1,19){$y$}
    \put(68,19){$y$}
    \put(23,1){$x$}
      \put(60,1){$t$}
        \put(90,1){$x$}
    \put(33,19.5){$F_\text{esc}$}
    \end{overpic}
    \caption{\label{fig:Fig3} 
    { Smart swimmers}. ({\bf a}) Intersection of phase-space LCS for constant steering ($\Xi = 0,\pm 0.2$), shown for $\theta = 0$, $\Xi=0.2$ (purple),  $\Xi=0$ (green), and $\Xi=-0.2$ (pink).
    ({\bf b}) $F_{\rm esc}(t)$ 
    for smart swimmers  with optimal time-dependent strategy (thick orange line), 
  for $\ve u=0$ [same as Fig.~\ref{fig:Fig2}({\bf a})],  black line, and for limited control ($\Xi = 0,\pm 0.2$), dashed orange line. ({\bf c}) Initial conditions that cannot escape with limited control (for $\theta=0$). 
    Other parameters: $\Phi = 0.1$, $B=0$.}
\end{figure}
The curves (red and blue solid lines) lie below $F_{\rm esc}(t)$ for $\ve u=0$ (black line), indicating that
the swimmers are trapped.
For steady flow ($B=0$, solid red line), $F_{\rm esc}(t)$ reaches a plateau below unity, showing that swimmers remain trapped ad infinitum. The plateau level, the fraction of trapped swimmers, is given by unity minus the fraction of two volumes: that of  the outermost  elliptic LCS $\gamma_\infty$ and that of the phase-space cell (red dashed line). Comparing the two curves, we see that most swimmers outside $\gamma_\infty$ have  escaped the vortex cell at $t\sim 20$. 
Since initial conditions inside $\gamma_\tau(t=0)$ also lie inside $\gamma_{\tau_1}(t=0)$ for $\tau_1<\tau$,  the phase-space volume of $\gamma_t(t)$ determines
 an upper bound on $F_{\rm esc}(t)$  (red, $\square$). Since  swimmers outside $\gamma_t(t)$ quickly leave the vortex cell,  this bound is close. 

The escape fraction for unsteady flow ($B=0.04$, blue line) is larger than for $B=0$, but lies below $F_{\rm esc}(t)$ for $\ve u=0$. In other words, there are still transport barriers, but they are weaker than in steady flow because  fluctuations in $\ve u(\ve r,t)$ shrink $\gamma_\tau(t)$ as $\tau$ increases, causing the escape fraction to grow [blue $\vartriangle$ in Fig.~\ref{fig:Fig2}({\bf a})]. This upper bound on $F_{\rm esc}$ is not as close as for $B=0$, because trajectories escaping $\gamma_\tau$ take a significant time to pass hyperbolic LCS of the system. This is explained in panels ({\bf b}-{\bf e}) which show intersections of 
repelling hyperbolic LCS with the planes $\theta=0$ and $\pi/2$  for steady and unsteady flow, $B=0$ and $0.04$, for $\tau=20$, red and blue solid lines (details in SM \cite{sm}).
The panels also show the initial positions of particles not escaped at $t=20$ (black dots).
The main difference between steady and unsteady flow lies in the tangle of the hyperbolic LCS. 
For $B=0$, they cross the vortex-cell boundaries,  enabling rapid escape: most particles starting outside $\gamma_{\tau=20}(0)$ have escaped at $t=20$ (so the upper bound in panel~({\bf a}) is close). 
For $B=0.04$, by contrast, escape is much slower because the  hyperbolic LCS partition phase space 
so that there are large regions of slow escape. Therefore the upper bound in panel ({\bf a}) is not close.
In summary, the outermost elliptic phase-space LCS  acts as a transport barrier for swimmers. 
In unsteady flow, the LCS  shrink in volume, yet escape from the vortex cell remains slow due barriers formed by hyperbolic LCS.

{\em Escape via stochastic reorientation}.
 Microswimmers tend to experience rotational diffusion, therefore one must consider how this affects their transport. \citet{khurana2012interactions} observed that swimmers with rotational diffusion tend to reside within elliptic flow LCS $\gamma_\tau^{({\rm flow})}$  for long times, 
before the swimmers eventually escape. This can be understood in terms of elliptic phase-space LCS which 
are barriers for weak diffusion~\cite{blazevski2014hyperbolic,haller2018material}, because they maximise local shear towards the LCS.
For isotropic diffusion in two phase-space dimensions, \citet{haller2018material}  proved rigorously that elliptic LCS minimise diffusive transport. Our problem is not isotropic, but we now show for $B=0$ how the tilt of the elliptic phase-space LCS $\gamma_\infty$
determines the diffusive escape rate $J$. Since advective transport is tangential to $\gamma_{\infty}$, the flux through $\gamma_\infty$ is purely diffusive.
Since  advection on $\gamma_\infty$ is ergodic in our case, and since weak diffusion is slower than advection,  the concentration $\varrho$ becomes uniform on  $\gamma_\infty$, and 
 the concentration gradient $\ve\nabla \varrho$ points along
the local outward surface normal $\ve t$ of $\gamma_\infty$,
$\ve\nabla \varrho= (\ve t \cdot\ve \nabla\varrho)\ve t$. This allows us to write
\begin{equation}
\label{eq:J}
J/A_\infty  = - D_\theta \frac{{\rm d}\varrho}{{\rm d}{\epsilon}}\sum_k
\Big\langle \Big |\frac{\partial E_\infty}{\partial \theta_k}\Big|\Big\rangle\,.
\end{equation}
Here $A_\infty$ is the surface area of $\gamma_\infty$, $D_\theta$ is the angular diffusion constant,  $E_\infty(x,y,\theta)=\epsilon$ is a parameterisation of $\gamma_\infty$, $\partial E_\infty/\partial \theta_k$ is evaluated at
the explicit parameterisation $\theta_k(x,y,\epsilon)$ for $k=1,2,\ldots$, and the average is over the projection of $\gamma_\infty$ onto the $x$-$y$-plane.
Eq.~(\ref{eq:J}) shows that $\gamma_\infty$ must tilt to produce a non-zero escape flux. 

\newpage 
{\em Smart turning}. 
As seen above, stochastic turning allows to escape from $\gamma_\infty$, albeit inefficiently for small $D_\theta$.
The search for more efficient deterministic escape strategies by reinforcement learning~\cite{colabrese2017flow,biferale2019zermelo,schneider2019optimal,alageshan2020machine,gunnarson2021Learning,qiu2022Navigation,xu2023long,calascibetta2023taming,mousavi2025short}
is constrained by the fact that 
shape and size of $\gamma_\infty$ depend on the deterministic swimming dynamics (\ref{eq:swim_dyn}). 
 For instance, constant turning with fixed $\Xi$ (positive or negative) introduces two new elliptic phase-space LCS [Fig.~\ref{fig:Fig3}({\bf a})] containing initial conditions that cannot escape the vortex cell by turning constantly with fixed $\Xi$  because $|\Xi|$ is too small. To find an escape strategy, we train swimmers to escape the vortex cell using $Q$-learning~\cite{Sutton2018,mehlig2021machine}.  At regular time steps, the swimmer measures its discretised phase-space position and responds by steering with an  optimal value of $\Xi$ for that step, either $0$  or $\pm 0.2$. See SM \cite{sm} for details. For an initial condition 
 inside all three LCS, the swimmer fails to  escape within $t/\tau_{\rm K} = 40$, because   a  phase-space dependent steering strategy is needed in this case, which is hard to find.
 Changing the initial orientation so that the swimmer starts outside the LCS for constant
 turning with $\Xi\!=\!-0.2$, the swimmer  learns to escape within $t/\tau_{\rm K} \sim 25$ by mainly turning with $\Xi=-0.2$. The found strategy is successful but not optimal (steering with constant $\Xi=-0.2$ allows to escape within $t/\tau_{\rm K} = 9.1$). 

For rapid turning (allowing $\Xi = 0$ or $\pm 1$),  $Q$-learning finds efficient escape strategies by simultaneously training swimmers with random initial orientations and positions in the vortex cell. 
The swimmers escape faster than for $\ve u=0$, indicating that they make use of the flow  to escape [Fig.~\ref{fig:Fig3}({\bf b})].
However, when the learned strategy is applied with limited control ($\Xi = 0,\pm 0.2$),
$F_{\rm esc}(t)$ reaches a plateau $\leq 1$, indicating a new trap. Panel ({\bf c}) shows  the initial conditions of swimmmers that cannot escape. They form a dissipative attractor, because
the deterministic dynamics
no longer conserves phase-space volume since  $\omega_{\rm s}$ depends on $\theta$ for the optimal strategy (as the $Q$ table shows, see SM~\cite{sm}). This allows for  dissipative attractors to appear in phase space.  In summary, analysing phase-space structures is key to exploiting the flow and avoiding traps: even for the same flow, different swimming parameters create different phase-space structures, demanding parameter-specific strategies.

{\em Discussion.}
Our computations demonstrate that phase-space LCS explain transient trapping of microorganisms navigating complex unsteady flow.
In our example, phase space is three-dimensional, and hyperbolic LCS appear as open surfaces, while elliptic LCS form tilted tubular surfaces.
These phase-space LCS explain a range of behaviours observed in numerics 
\cite{khurana2011reduced,khurana2012interactions,torney2007transport,qin2022confinement}.
Since the actual phase-space barriers differ from the flow LCS, swimmers  may penetrate  flow LCS, as observed in Refs.~\cite{khurana2011reduced,khurana2012interactions}. Since the tilt of the phase-space LCS is not uniform,   different swimming orientations give rise to different degrees of barrier penetration depending on swimmer orientation [Fig.~\ref{fig:Fig1}({\bf c})].
Second, transient trapping of slow swimmers ($\Phi\ll 1$) outside flow traps~\cite{khurana2011reduced} can be explained by hyperbolic LCS.
Third,   \citet{qin2022confinement} found that swimmers 
in time-periodic flow are confined to flow manifolds for small $\Phi$. For their parameters, the elliptic LCS depends only weakly on $\theta$. In this special case, the flow manifold is a good approximation. For larger $\Phi$ it is not, as seen in Fig.~S3 in the SM~\cite{sm}.
Fourth, while numerical simulations~\cite{torney2007transport} indicate long-term trapping in unsteady flow for parameter values similar to those in Fig.~\ref{fig:Fig2}, our results show that 
trapping is  transient. The  fraction  $F_{\rm esc}(t)$ increases monotonically to unity [Fig.~\ref{fig:Fig2}({\bf a})], allowing swimmers to eventually escape, and indicating that the elliptic phase-space LCS shrinks until it disappears.

Regarding our choice of parameter values, we note that ocean turbulence spans scales from mesoscale biomass transport~\cite{froyland2015studying} to small eddies that confine individual microorganisms.
Fast microswimmers escape small traps but not large ones. At intermediate scales, escape depends on swimming behavior~\cite{torney2007transport,durham2011gyrotactic,khurana2011reduced,arguedasleiva2020microswimmers,qin2022confinement,tanasijevic2022microswimmers}.
The velocity of turbulent fluctuations, $u_{\rm rms}$, increases with the dissipation rate $\varepsilon$~\cite{yamazaki1996comparison}.
We considered $v_{\rm s}\sim 0.1u_{\rm rms}$, representing weak swimmers such as {\em G. dorsum} in mid-depths ($\varepsilon \sim $ \SI{0.1}{\milli\metre\squared\per\s\cubed}), or stronger swimmers such as {\em M. pacifica} near the ocean surface ($\varepsilon\sim $ \SI{20}{\milli\metre\squared\per\s\cubed})~\cite{yamazaki1996comparison}.

Our results generalise to other systems, but
we note that the types of LCS depend on the system in question.
Orientation-dependent steering (or shape
asymmetries) lead to dissipative dynamics, 
allowing other finite-time phase-space structures: attractors or repellers, including isolated elliptic LCS (analogous to limit cycles), or globally attracting or repelling hyperbolic LCS~\cite{haller2013coherent,beneitez2020edge}. 

Also in higher dimensions, 
LCS act as transport barriers and traps, although they become harder to construct.
For a swimmer in three spatial dimensions, for example, phase space is five-dimensional, requiring to identify  four-dimensional material surfaces: ridges of maximal FTLE (hyperbolic LCS) or uniformly stretching surfaces (elliptic LCS). We expect that the latter form along vortex tubes and are transported along the tube.

Reinforcement learning produces strategies allowing swimmers to navigate barriers shaped by their own behavior: changing the behavior changes the barrier landscape.
For deterministic steering, 
LCS depend on strategy.  
In Ref.~\cite{krishna2023finite}, hyperbolic LCS based on a control strategy were used to analyse swimming behavior.
Since reorientation is instantaneous in Ref.~\cite{krishna2023finite},
these LCS reflect effective flow LCS rather than the phase-space barriers studied here.
In general, strategy-dependent LCS may  guide navigation across a wide range of flow configurations. 
However,  as flow statistics change, a trained strategy may become suboptimal.
Efficient navigation thus relies on detecting and adapting to the phase-space structures associated with different strategies and flows.

{\em Conclusions.}
Understanding traps and barriers for active microorganisms in complex flow requires analysis in phase space, combining both hydrodynamic flow transport and swimming dynamics.
In vortex flow, hyperbolic and elliptic phase-space LCS explain transient trapping. In steady flow, elliptic LCS may trap indefinitely, though their tilt enables escape via reorientation.
Stochastic reorientation leads to slow escape, dominated by regions of large LCS tilt,  while reinforcement-learning strategies enable rapid escape when turning rates are sufficient.

Most reinforcement-learning approaches rely on local, instantaneous signals~\cite{colabrese2017flow,biferale2019zermelo,schneider2019optimal,alageshan2020machine,gunnarson2021Learning,qiu2022Navigation,xu2023long,calascibetta2023taming,mousavi2025short},
which may miss global flow structures.
Our results suggest that optimal strategies are constrained by non-local phase-space structures determined by both flow and swimming behaviour.
Efficiently predicting and exploiting transport barriers therefore requires non-local flow information, for example by sampling flow histories, or by communication in a network of swimmers.
We hypothesize that many efficient navigation strategies remain to be discovered, particularly those that take advantage of global flow features not accessible by local signals.

{\em Acknowledgments.}
We acknowledge support from the Knut and Alice Wallenberg Foundation, grant no. 2019.0079, and from Vetenskapsr\aa{}det,  grant nos. 2018-03974, 2023-03617 (JQ and KG), and 2021-4452 (LS, BM). Numerical computations were performed on resources provided by the Swedish National Infrastructure for Computing (SNIC), partially funded
by the Swedish Research Council through Grant Agreement No. 2018-05973.

\bibliography{references}
\end{document}


\title{Supplemental material for \lq Transport barriers for microswimmers in unsteady flow\rq{}}
\author{L. Storm}
\affiliation{Department of Physics, University of Gothenburg, 41296 Gothenburg, Sweden}
\author{J. Qiu}
\affiliation{Department of Physics, University of Gothenburg, 41296 Gothenburg, Sweden}
\author{K. Gustavsson}
\affiliation{Department of Physics, University of Gothenburg, 41296 Gothenburg, Sweden}
\author{B. Mehlig}
\affiliation{Department of Physics, University of Gothenburg, 41296 Gothenburg, Sweden}

\begin{abstract}
In this document, we summarise details regarding the 
definitions of elliptic and hyperbolic LCS (Section \ref{sec:defs}), and how to find them (Sections \ref{sec:detection} to IV). Details 
regarding the reinforcement learning are described in
Section~\ref{sec:rl}, and Section~\ref{sec:suppdata} contains supplemental data.
\end{abstract}

\maketitle

\setcounter{figure}{0}
\setcounter{table}{0}
\makeatletter
\renewcommand{\thefigure}{S\arabic{figure}}
\renewcommand{\thetable}{S\arabic{table}}
\renewcommand{\theequation}{S\arabic{equation}}

\onecolumngrid

\section{Definitions of LCS}
\label{sec:defs}
In order to document the numerical methods used to obtain Figs.~1 and 2 in the main text as well as  Fig.~S3, we briefly  review  the relevant mathematical definitions concerning Lagrangian coherent structures (LCS) from Refs.~\cite{oettinger2016global,oettinger2016autonomous,blazevski2014hyperbolic,haller2011variational}. 
\subsection{Deformation tensor}
The equation of motion (1) in the main text is a first-order autonomous dynamical system of the form
\begin{equation}
\label{eq:flow}
\tfrac{\rm d}{{\rm d}\tau}\ve y_t 
= \ve f(\ve y_t,t)\,.
\end{equation}
Here $\ve y_t = [\ve r_t,\theta_t]$ denotes the phase-space point at time $t$ given the initial condition $\ve y_0$. An infinitesimal perturbation
$\delta \ve y_0$ of $\ve y_0$ is mapped as 
\begin{equation}
   \delta  \ve y_t=\ma J_t \,\delta \ve y_0\,,
    \end{equation}
    where $\ma J_t$ is the deformation
    tensor or monodromy matrix \cite{wilkinson2009fingerprints,meibohm2020fractal} with elements $
J_{tij} = \partial y_{ti}/\partial y_{0j}$.
The deformation tensor  obeys the differential equation
\begin{equation}
    \label{eq:deform}
    \tfrac{\rm d}{{\rm d}t}
    {\ma J}_t=\ma W\big(\ve y_t,t\big)\ma J_t\,, \quad\mbox{with} \quad\ma J_0=
    \ma I\,.
\end{equation}
Here $\ma W$ is the linearisation of the phase-space flow (\ref{eq:flow}), with elements
 $W_{ij} = \partial f_i/\partial y_j$, and $\ma I$ is the unit matrix. 

\subsection{Elliptic LCS}
We use different definitions for elliptic LCS of the flow (two-dimensional, 2D), and elliptic phase-space LCS (three-dimensional, 3D). 

In two dimensions,
we start from Ref.~\cite{haller2013coherent}, where an elliptic LCS  $\gamma_\tau$ is defined as a closed contour of initial conditions that deforms with the same rate at every point on the contour, over time $\tau$.  It is shown in Ref.~\cite{haller2013coherent} that such a contour must be {\em tangential} to the vector field
\begin{eqnarray}
    \label{eq:elliptic_LCS_2D}
    \ve \eta^{(\lambda)}_\pm (\ve y_0)=\sqrt{\frac{\sigma_1^2-\lambda^2}{\sigma_1^2-\sigma_2^2}}\ve v_2\pm\sqrt{\frac{\lambda^2-\sigma_2^2}{\sigma_1^2-\sigma_2^2}}\ve v_1\,.
\end{eqnarray}
Here $\sigma_1\geq\sigma_2\geq\dots$  are 
the singular values of the deformation tensor $\ma F_\tau$ corresponding to the phase-space trajectory from $\ve y_0$ to $\ve y_\tau$,  and  $\ve u_\alpha$ and $\ve v_\alpha$ are the corresponding left and right singular vectors.
 Such a contour has the property that every infinitesimal line segment on it is stretched by the factor $\lambda>0$ over the integration time $\tau$. 

In three dimensions, 
an elliptic LCS is a material surface that experiences point-wise uniform stretching in its tangent plane over the time horizon $\tau$. Such surfaces are  {\em normal} to the vector field \cite{oettinger2016global}
\begin{equation}
    \label{eq:eta_field}
    \boldsymbol{\eta}_\pm(\ve y_0)=\sqrt{\frac{\sigma_2^2-\sigma_3^2}{\sigma_1^2-\sigma_3^2}}\ve v_3\pm\sqrt{\frac{\sigma_1^2-\sigma_2^2}{\sigma_1^2-\sigma_3^2}}\ve v_1\,.
\end{equation}
Tangential stretching equals the second singular value $\sigma_2$ of the deformation tensor. Since material surfaces that are precisely uniformly stretching are rare, \citet{oettinger2016global} define an almost-uniformly stretching material surface by allowing small deviations from the normal vectors in Eq.~(\ref{eq:eta_field}) using the vector field
\begin{equation}
    \label{eq:eta_field_dev}
    \boldsymbol{\eta}^{(\delta)}_\pm(\ve y_0)=\sqrt{\frac{\sigma_2^2(1+\delta)-\sigma_3^2}{\sigma_1^2-\sigma_3^2}}\ve v_3\pm\sqrt{\frac{\sigma_1^2-\sigma_2^2(1+\delta)}{\sigma_1^2-\sigma_3^2}}\ve v_1,
\end{equation}
where $\sigma_1>\sigma_2\sqrt{1+\delta}>\sigma_3$. We use Eq.~(\ref{eq:eta_field_dev}) for our three-dimensional numerical computations.

\subsection{Hyperbolic LCS}
A hyperbolic LCS is a finite-time generalisation of stable and unstable manifolds \cite{haller2011variational}. The definition is the same in 2D and 3D. A repelling hyperbolic LCS is a material surface that locally normally repels nearby trajectories the most over the time horizon $\tau$. Attracting hyperbolic LCS are defined in the converse way. In \cite{haller2011variational} it is shown that such a material surface must be orthogonal to the maximal (repelling) or minimal (attracting) right singular vector of the deformation tensor. This means that repelling (attracting) hyperbolic LCS coincides with ridges of maximal (minimal) finite-time Lyapunov exponents (FTLE) \cite{haller2011variational}.

\section{Finding elliptic LCS}\label{sec:detection}
\subsection{Two-dimensional phase space}
In order to find elliptic LCS in 2D phase space, we integrate Eq.~(\ref{eq:elliptic_LCS_2D}) for different values of $\lambda$ and find a nested family of elliptic LCS. In the main text, we denote the outermost of these by $\gamma_\tau^{({\rm flow})}$.

\subsection{Three-dimensional phase space}
To find elliptic LCS in 3D phase space, we proceed in two steps. First, we locate a collection of points on the  LCS. Second, we integrate a trajectory starting at these points that samples the LCS, using the method of Ref.~\cite{oettinger2016autonomous}.

To find initial points on the LCS, we make use of the method introduced by \citet{blazevski2014hyperbolic}. They introduce a 
horizontal plane at height $y_3 = y_3^\ast$ with normal vector $\ve n_\ast$, such that it passes through a potential LCS. In this case, the vector $\ve v_\parallel=\ve n_\ast\wedge\boldsymbol{\eta}^{(\delta)}_\pm$ is tangential to the elliptic  LCS.
An intersection of an elliptic LCS is therefore a trajectory of the dynamical system
\begin{eqnarray}
    \label{eq:contour}
    \dot{\ve r}=\ve v_\parallel(\ve r).
\end{eqnarray}
In  Ref.~\cite{blazevski2014hyperbolic}, it is shown that for a contour of Eq.~(\ref{eq:contour}) to be an intersection of an LCS, it must have zero helicity $[\nabla\wedge\boldsymbol{\eta}_\pm(\ve y_0)]\cdot \boldsymbol{\eta}_\pm(\ve y_0)$ on every point of the contour. We use this to verify that the detected contours belong to elliptic LCS.

To find closed contours in the dynamical system in Eq.~(\ref{eq:contour}), we start by computing maximal FTLE over a grid of $400\times400$ points in the $y_1$-$y_2$-plane for a fixed valueod $y_3$. This guides the search for elliptic LCS, as their interiors experience only weak deformation, and therefore form islands of small maximal FTLE values. The dynamics in Eq.~(\ref{eq:contour}) is then integrated starting from a grid of 10 points along the $y_1$-axis from $-\pi/2$ to $\pi/2$, and with $y_2$ selected to lie at the center of the island of small maximal FLTE values, to make a rough estimate of the location of the LCS. The $y_1$-grid is then refined to find the outermost elliptic contour with a precision of $\delta y_1=\pm0.01$. We seek contours with an average helicity no greater than $0.01$.

The second step
 relies on the fact that, as shown in Ref.~\cite{oettinger2016autonomous}, hyperbolic and elliptic LCS are invariant structures of the dynamical
\begin{eqnarray}
    \label{eq:v2_dyn}
    \dot{\ve y}_0=\ve v_2(\ve y_0)\,.
\end{eqnarray}
Starting a trajectory on an LCS, the dynamics given by Eq.~(\ref{eq:v2_dyn}) stays on the LCS.  We note that elliptic LCS are closed quasi-periodic structures of the $\ve v_2$-dynamics, whereas hyperbolic LCS are open sheets. 

Once a closed contour has been obtained, Eq.~(\ref{eq:v2_dyn}) is integrated starting from 100 evenly spaced points on the contour. We integrate until the trajectories trace out a dense surface. To verify that the surface is an elliptic LCS, we compute the normal vectors of the material surface and compare them to $\boldsymbol{\eta}^{(\delta)}_\pm(\ve y_0)$. To compute the normal vectors of the material surface, we use the Python package \texttt{open3d} \cite{zhou2018open3d} to construct a triangle mesh from the point cloud obtained by integrating along the dynamics in Eq.~(\ref{eq:v2_dyn}). The triangle mesh is constructed using screened Poisson reconstruction~\cite{kazhdan2006poisson}. Once the triangle mesh has been obtained, the vector obtained from Eq.~(\ref{eq:eta_field_dev}) is computed at the center of each triangle mesh and compared to the normal vector of the triangle by computing their cosine similarity. We compute the average cosine similarity, where for every triangle, we find the value of $\delta$ with the highest cosine similarity, subject
to $-0.2\leq \delta\leq 0.2$.

The elliptic LCS in Figure~1 in the main text are computed for $\Phi=0.1$ in unsteady  ($B=0.04$, $\omega=0.5$) flow for $\tau=5$ and $\tau=10$. To obtain the evolved material surface, we uniformly sample $10^5$ points on the triangle mesh and evolve them according to the swimmer dynamics in Eq.~(1) in the main text. A triangle mesh is then obtained for the evolved point clouds. In panel (c), a triangle mesh could not be constructed for $\gamma_{\tau=5}(10)$ due to the significant deformation of the material surface. The average cosine similarity for the detected elliptic LCS are $0.998$ and $0.9902$ for $\tau=5$ and $\tau=10$ respectively, averaged over $5\times10^4$ points on the mesh. The 2D elliptic LCS are computed for $\Phi=0.1$ in an unsteady  flow ($B=0.04$, $\omega=0.5$) for $\tau=5$ and $\tau=10$. We sweep $\lambda$ over 10 evenly spaced points in the range $[0.8,1.2]$ and find that $\lambda=1\pm0.04$ for both values of $\tau$.

The elliptic LCS in Figure~2 in the main text are computed for $\Phi=0.1$ in unsteady  flow ($B=0.04$, $\omega=0.5$). The volume contained in the elliptic LCS is computed using an in-built method in the \texttt{open3d} module. The average cosine similarity is above $0.99$ for all LCS, averaged over $5\times10^4$ points on the mesh.

\section{Finding hyperbolic LCS}
To find initial points on hyperblic LCS,
we can use (\ref{eq:contour}), but now with 
$\ve v_\parallel$ equal to 
 $\ve v_1$ for repelling hyperbolic LCS, and $\ve v_3$ for attracting hyperbolic LCS. 
To guide the search, we compute maximal FTLE over a grid of $400\times400$ points in the $y_1$-$y_2$ plane for a fixed  value of $y_3$. To verify whether FTLE ridges correspond to intersections of repelling hyperbolic LCS, we also compute the helicity $[\nabla\wedge\ve v_1(\ve y_0)]\cdot \ve v_1(\ve y_0)$ over a grid of $400\times400$ points. We find that the helicity is smaller than $10^{-8}$ along the ridges of maximal FTLE shown in Fig.~2({\bf b}-{\bf e}), confirming that they are intersections of repelling hyperbolic LCS. We also find that trajectories of Eq.~(\ref{eq:contour}) launched from the ridges of maximal FTLE follow the ridges (not shown).

\section{Integrating $\ma J_\tau$}
Equation~(\ref{eq:deform}) may be numerically unstable if integrated for several Lyapunov times. The numerical instability can be circumvented by instead integrating left Cauchy-Green tensor $\ma B_t=\ma J_t\ma J_t^{\sf T}$ by decomposing it into rotational and stretching dynamics~\cite{balkovsky1999universal,meibohm2020fractal}. Since we are interested in the eigenvectors of the right Cauchy-Green tensor $\ma C_t=\ma J_t^{\sf T}\ma J_t$, which are equal to the right singular vectors of $\ma J_t$, we make use of the fact that the right Cauchy-Green tensor computed at $\ve y_0$ is $\ma C_t=(\ma B_{-t})^{-1}$, where $\ma B_{-t}$ is computed by integrating the dynamics backwards in time from the final position $\ve y_t$ to the starting position $\ve y_0$. 

\section{Reinforcement learning}\label{sec:rl} 
We use $Q$-learning~\cite{Sutton2018,mehlig2021machine} to find smart turning strategies allowing swimmers to escape the vortex cell (see main text). 
The training is divided into episodes labelled by the integer $k=1,\ldots,K$, where
$K$ is the total number of episodes. 
At regular times $t = n\Delta t$ during each episode (where $n=0,1,2,\ldots,N$ and $N$ is the total number of steps per episode), the swimmer measures its state $\ve s_n$, namely its discretised phase-space position $\ve y_{t}=[\ve r_{t},\theta_{t}]^{\sf T}$ on a uniform $16^3$ grid. Given the state $\ve s_n$, the swimmer reacts by choosing an action $a_n$: it either turns clockwise ($\Xi>0$),  anticlockwise ($\Xi<0$), or it swims straight ahead ($\Xi=0$). Here $\Xi = \omega_s/\tau_{\rm K}$ is the non-dimensional angular velocity (see main text). 
Which of these actions is taken is determined using an epsilon-greedy policy \cite{mehlig2021machine}. The chosen action determines the new state $\ve s_{n+1}$ and the instantaneous reward:
\begin{equation}
    R_{n+1} = \left\{\begin{array}{r@{\quad}l}
-0.005 & \text{if $\ve r_{n+1}$ is inside the vortex cell}, \\
1 & \text{otherwise}.
\end{array}
\right.
\end{equation}
We see that the swimmer receives the $R_{n+1}=1$ if $\ve r_{n+1}$ is from the vortex cell. 
Otherwise, the swimmer is penalised for being trapped. The penalty is chosen as $N^{-1}=0.005$, where $N=2000$ is the total number of state updates per episode and $N\Delta t$
is approximately 10 times longer than the typical escape time without flow. 

Optimal strategies are found by iteratively updating the elements of  the $Q$-table $Q(\ve s,a)$,  using $Q(\ve s_n,\ve a_n)\leftarrow Q(\ve s_n,\ve a_n) + \alpha [R_{n+1} + \gamma \max_a Q(\ve s_{n+1},a)-Q(\ve s_n,a_n)]$, where $\alpha$ is the learning rate, and $\gamma=0.999$ is the discount rate, which is chosen close to one to value long-term rewards over a time horizon of $(1-\gamma)^{-1}\Delta t=10^3\Delta t$.
The elements of the Q-table are initialised using the optimistic initial value 0.8 to encourage exploration, yet small enough to ensure training efficiency.
In each episode, one hundred swimmers are initialised uniformly in a square region with certain initial orientations, then the dynamics of the swimmers are integrated with $\delta t/\tau_{\rm K}=0.001$. A swimmer is removed from the simulation once it has escaped from the vortex cell. The episode ends if the maximum episode length is reached, or if all swimmers have escaped. 
During training, the learning rate decays with the number of episode $k$ using $\alpha = \alpha_0 k_0/(k_0 + k)$, where the initial learning rate $\alpha_0=0.05$, and $k_0=8000$. The probability of choosing a random action  decays as  $\epsilon = \epsilon_0 \max (0, 1-k/\todo{k_0})$, where $\epsilon_0=0.01$.

\begin{figure}
    \centering
    \includegraphics[width=1\linewidth]{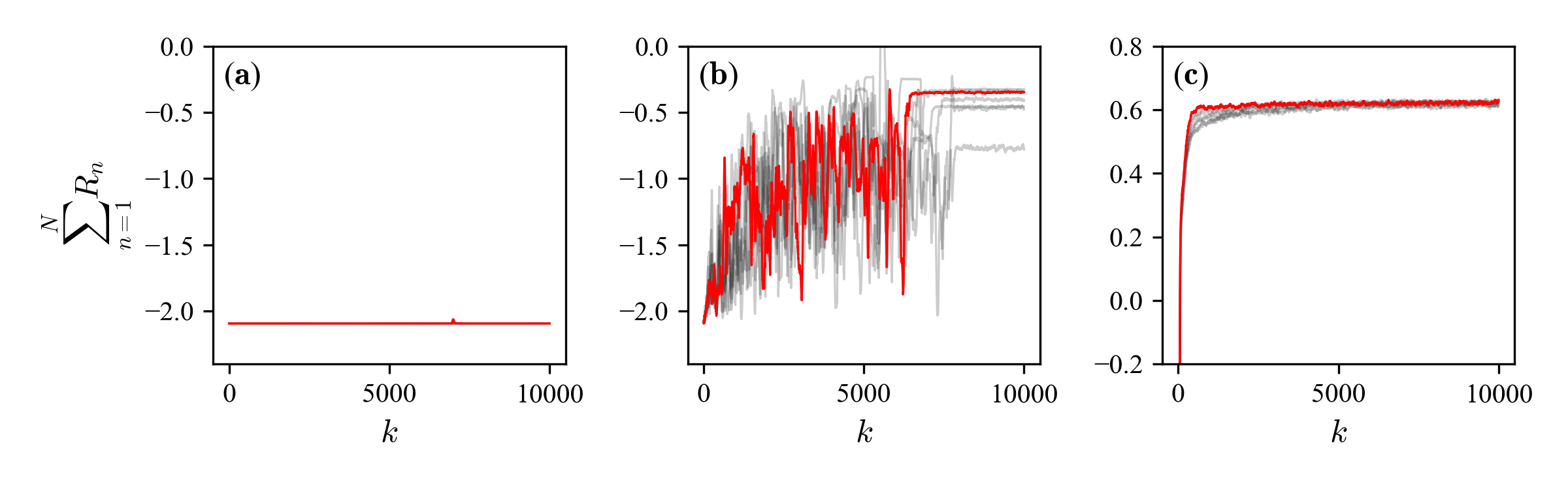}
    \caption{Training curves show the total reward in one episode $\sum_{n=1}^{N} R_n$ as a function of the number $k$ of episode. 
    Red curves: training histories of the strategies that yield the largest total reward. The other training histories are shown as gray curves. All data is smoothed using a sliding average over a window of $50$ episodes. 
    ({\bf a}) $\Xi=0.2$, $N =400$, and swimmers are initialised uniformly in a small region, $0.96 \leq x\leq 1.04 $ and $0.04 \leq y\leq -0.04 $, with initial orientation $\theta = \pi/2$, 
    which are inside all three LCS. ({\bf b}) Same as ({\bf a}) but with initial orientation $\theta =0$, which is outside the $\Xi=-0.2$-LCS. 
    ({\bf c}) $\Xi = 1$, $N =2000$, and swimmers are initialised uniformly in the vortex cell with random orientations.
    \label{fig:figS2}}
\end{figure}

In the main text, we discuss three different cases with different values of $\Xi$ and different initial conditions. For each case, training are repeated seven times using different random seeds, and we choose the strategy that yields the largest total reward.
For smart swimmers with $\Xi=\pm0.2$ or 0 (see main text), the swimmers are initialised uniformly in $0.96 \leq x\leq 1.04 $ and $0.04 \leq y\leq -0.04$, with two initial orientations, $\theta=\pi/2$ for swimmers inside all three LCS, and  $\theta = 0$ for swimmers out side the $\Xi=-0.2$-LCS. 
The total number of state updates per episode is $N=400$. The corresponding training curves are shown in Fig.~\ref{fig:figS2}({\bf a}, {\bf b}). Swimmers inside all three LCS do not learn to escape
 [panel ({\bf a})]. Swimmers
 outside the $\Xi=-0.2$-LCS learn to escape: the total reward  $\sum_{n=1}^{N} R_n$ increases with the number $k$ of training episodes, and reaches a plateau indicating that the strategy converges.
 
\begin{figure}
    \centering
    \includegraphics[width=0.8\linewidth]{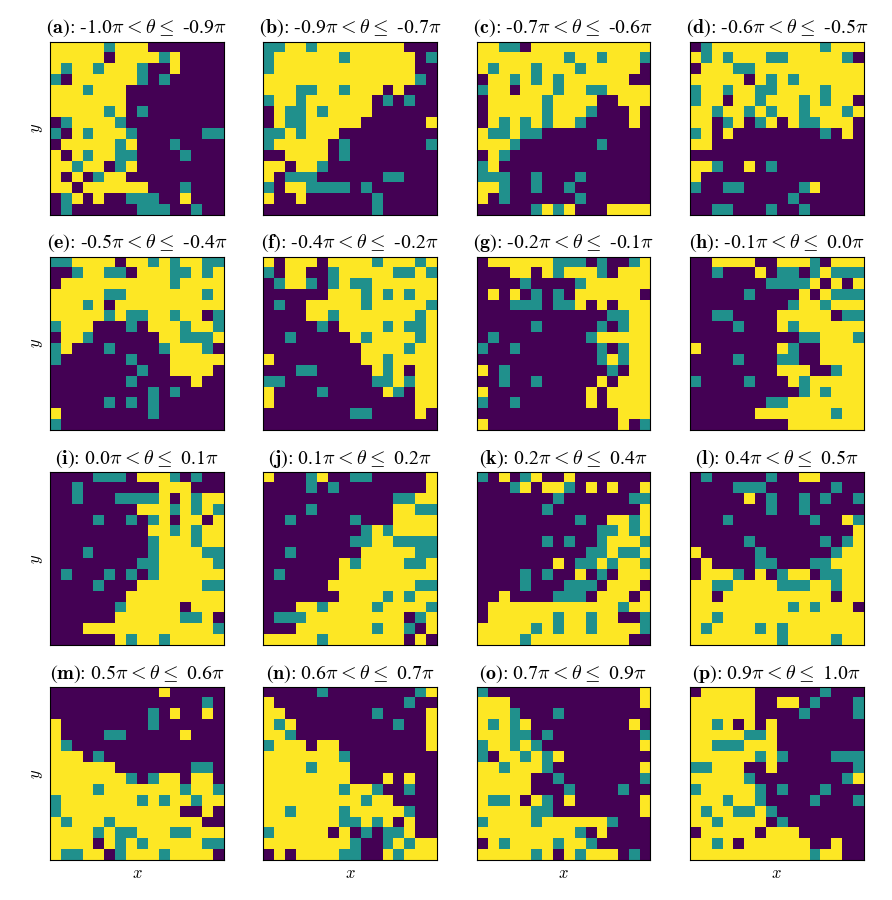}
    \caption{Q-table trained for $\Xi = 1$, corresponding to the orange curves in Fig.~3 in the main text.  
    Each panel shows the optimal actions $\mbox{argmax}_a Q(\ve s,a)$ for different values of $\ve r$ and $\theta$. Yellow: $\Xi=1$. Blue: $\Xi=-1$. Green: $\Xi=0$. }
    \label{fig:figs3}
\end{figure}

For rapidly-turning swimmers (Fig.~3 in the main text, $\Xi=\pm1$ or 0), the swimmers are initialised uniformly in the vortex cell, $-\pi /2\leq x, y\leq \pi /2$ with random orientations, identical to non-smart swimmers discussed in Fig.~2 in the main text. The total number of state updates per episode is $N=2000$. The training curves are shown in Fig.~\ref{fig:figS2}({\bf c}). The $Q$-table corresponding to the best strategy found in the trainings is documented in Fig.~\ref{fig:figs3}.

\section{Supplemental data}\label{sec:suppdata}
\begin{figure}
    \centering
    \begin{overpic}[width=18cm]{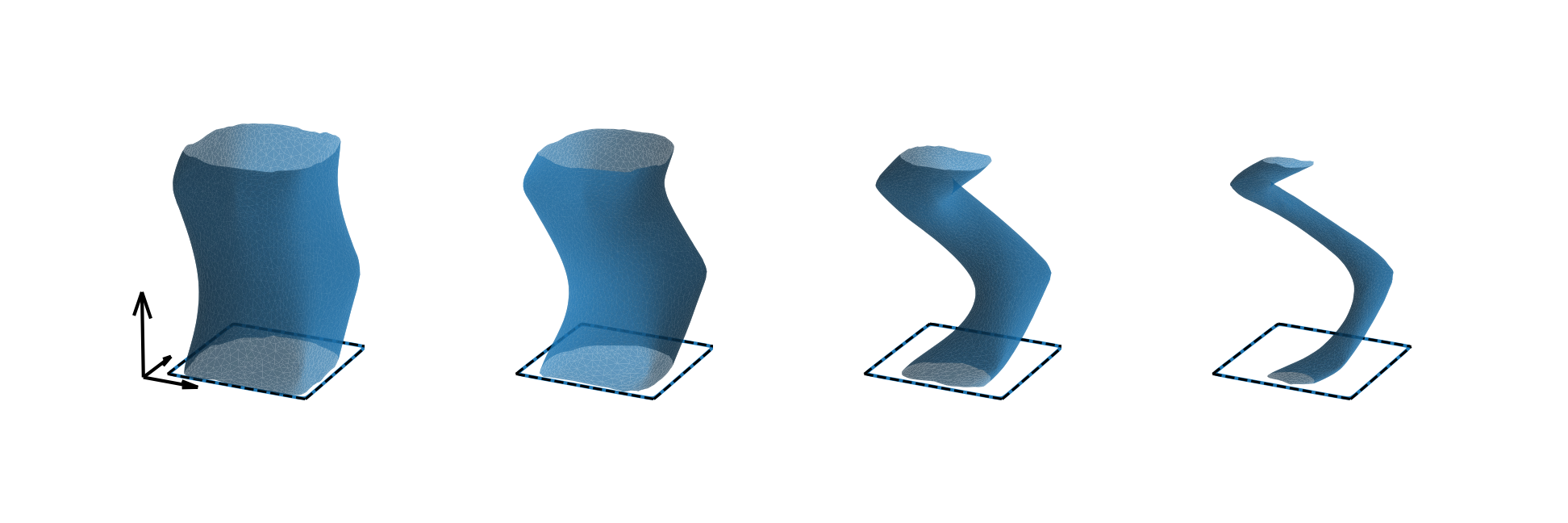}
    \put(13,3){$\Phi=0.05$}
    \put(36,3){$\Phi=0.1$}
    \put(59,3){$\Phi=0.2$}
    \put(82,3){$\Phi=0.3$}
    \put(13,7.5){$x$}
    \put(8.8,15.3){$\theta$}
    \put(10.5,11.5){$y$}
    \put(8,25){({\bf a})}
    \put(31,25){({\bf b})}
    \put(54,25){({\bf c})}
    \put(77,25){({\bf d})}
    \end{overpic}
    \caption{
    Outermost elliptic LCS 
      (Lagrangian vortex boundary) $\gamma_{\infty}$ for different values of $\Phi$. 
    Parameters $\omega_{\rm s}=B=0$. The blue line shows the boundary of the vortex cell..}
    \label{fig:Fig1_supp}
\end{figure}
In Fig.~\ref{fig:Fig1_supp}, the outermost elliptic LCS (Lagrangian vortex boundary) $\gamma_\infty$ for steady flow are shown for different swimming speeds $\Phi$. See Section~\ref{sec:detection} for details on the simulation. To get a good approximation of the infinite-time structures, we use $\tau=200$, although the structures converge before $\tau=50$.

\bibliography{references}